\documentclass[showpacs,superscriptaddress,preprintnumbers,floats,eqsecnum,pra,balancelastpage,onecolumn]{revtex4}
\usepackage{amssymb}
\usepackage{amsfonts}
\usepackage{amsmath}
\usepackage{hyperref}
\usepackage{scalefnt}
\usepackage{graphicx}%
\newtheorem{theorem}{Theorem}
\newtheorem{conjecture}[theorem]{Conjecture}

\begin{document}
\title{Some families of density matrices for which separability is easily tested}
\author{Samuel L. Braunstein}
\affiliation{University of York, Heslington, York YO10 5DD, U.K.}
\author{Sibasish Ghosh}
\affiliation{University of York, Heslington, York YO10 5DD, U.K.}
\author{Toufik Mansour}
\affiliation{University of Haifa, 31905 Haifa, Israel.}
\author{Simone Severini}
\affiliation{University of York, Heslington, York YO10 5DD, U.K.}
\author{Richard C. Wilson}
\affiliation{University of York, Heslington, York YO10 5DD, U.K.}
\pacs{03.67.-a, 03.67.Mn}

\vspace{0.4cm}
\begin{center}
(Accepted for publication in {\it Phys. Rev. A})
\end{center}

\vspace{0.4cm}

\begin{abstract}
We reconsider density matrices of graphs as defined in
[quant-ph/0406165]. The density matrix of a graph is the
combinatorial laplacian of the graph normalized to have unit
trace. We describe a simple combinatorial condition (the
\textquotedblleft degree condition\textquotedblright) to test
separability of density matrices of graphs. The condition is
directly related to the PPT-criterion. We prove that the degree
condition is necessary for separability and we conjecture that it
is also sufficient. We prove special cases of the conjecture
involving nearest point graphs and perfect matchings. We observe
that the degree condition appears to have value beyond density
matrices of graphs. In fact, we point out that circulant density
matrices and other matrices constructed from groups always satisfy
the condition and indeed are separable with respect to any split.
The paper isolates a number of problems and delineates further
generalizations.

\end{abstract}
\maketitle

\section{Introduction}

Let us consider the set $\mathcal{S}$ of all density matrices
$\rho$ of a bipartite system with assigned Hilbert space
$\mathbb{C}_{A}^{p}\otimes$ $\mathbb{C}_{B}^{q}$. Let
$\mathcal{M}(pq)$ be the linear space (over the complex field
$\mathbb{C}$) of all $pq\times pq$ complex matrices equipped with
the inner product $\langle A|B\rangle:=$ Tr$\left(
A^{\dagger}B\right) $ for any $A,B\in\mathcal{M}(pq)$. Let us
consider the metric $D(A,B):=\langle (A-B)|(A-B)\rangle$ on
$\mathcal{M}(pq)$, for any $A,B\in\mathcal{M}(pq)$. With respect
to this metric, the set $\mathcal{S}$ forms a compact (which is
also convex) subset of $\mathcal{M}(pq)$ generated by $\left(  p^{2}%
q^{2}-1\right)  $ real parameters. Deciding whether a given
element from this compact set $\mathcal{S}$ is separable or
entangled (the \emph{separability problem}) is known to be NP-hard
\cite{d}. This problem is an instance of the weak membership
problem as defined by Gr\"{o}tschel \emph{et al.} \cite{gls} (see
also \cite{gv}). Recently the separability problem has been
considered and discussed in \cite{e, d, hb, io}.

There are few cases where the separability problem is known to be
efficiently solvable. The best known situation is $(p,q)=(2,3)$ or
$(3,2)$. In this case, the positivity of ${\rho}^{\Gamma_{B}}$
(the partial transposition of ${\rho}$ with respect to the system
$B$) is equivalent to the separability of $\rho$ \cite{P,H}. Also,
the set of all density matrices \textquotedblleft very
near\textquotedblright\ (in the sense of some useful metric) to
the maximally mixed state is known to be separable \cite{sam}.
Other examples are given in \cite{l}.

For some discrete family of density matrices (that is, for which
no continuity argument can be applied), no such efficient
criterion is known to us. Here we consider the family of the
\emph{density matrices of graphs} as introduced in \cite{b}. In
the remainder of this section we introduce some terminology and
state the results. There are two other sections in the paper:\
Section II contains the proofs; Section III\ is a list of further
problems and generalizations.

Let $G=(V,E)$ be a simple graph on $n$ labeled vertices, that is
$V=\{v_{1},v_{2},...,v_{n}\}$ and $E\subseteq V^{2}=\{\{v_{i},v_{j}%
\}:v_{i},v_{j}\in V$ and $i\neq j\}$. The \emph{adjacency matrix}
of $G$ is an $n\times n$ matrix, denoted by $M(G)$, with lines
indexed by the vertices of $G$ and $ij$-th entry defined as
\[
\left[  M(G)\right]  _{i,j}=\left\{
\begin{tabular}
[c]{cc}%
$1,$ & if $\{v_{i},v_{j}\}\in E(G);$\\
$0,$ & if $\{v_{i},v_{j}\}\notin E(G).$%
\end{tabular}
\ \right.
\]
The \emph{degree matrix} of $G$ is an $n\times n$ matrix, denoted
by
$\Delta(G)$, with $ij$-th entry defined as%
\[
\left[  \Delta(G)\right]  _{i,j}=\left\{
\begin{tabular}
[c]{ll}%
$|\{v_{j}:\{v_{i},v_{j}\}\in E\}|,$ & if $i=j;$\\
$0,$ & if $i\neq j.$%
\end{tabular}
\ \right.
\]
The \emph{laplacian} \cite{gr} of $G$ is the symmetric positive
semidefinite matrix
\[
L(G):=\Delta(G)-M(G).
\]
Other than this combinatorial laplacian, there are several other
types of
laplacians associated to graphs \cite{ch}. The matrix%
\[
\rho(G):=\frac{1}{2|E|}L(G)
\]
is a density matrix. This is called the \emph{density matrix of
}$G$ \cite{b}.

It should be noted here that the notion of density matrix
$\rho(G)$ of a graph $G = (V(G), E(G))$, as defined above, is
completely different from the notion of `graph states', introduced
by Briegel and Raussendorf \cite{br}. A graph state $|G\rangle$,
corresponding to a (simple) graph $G = (V(G), E(G))$ is a common
eigen state (corresponding to the eigen value $1$) of the $n =
|V(G)|$ no. of $n$-qubit operators ${\sigma}_{11} \otimes
{\sigma}_{12} \otimes \ldots \otimes {\sigma}_{1n}$,
${\sigma}_{21} \otimes {\sigma}_{22} \otimes \ldots \otimes
{\sigma}_{2n}$, $\ldots$, ${\sigma}_{n1} \otimes {\sigma}_{n2}
\otimes \ldots \otimes {\sigma}_{nn}$, where (i) ${\sigma}_{ii} =
{\sigma}_x$ for all $i \in \{1, 2, \ldots, n\}$, (ii) for $j \ne
i$, ${\sigma}_{ij} = {\sigma}_z$ if the vertices $v_i$ and $v_j$
of $G$ are connected by an edge, and (iii) for $j \ne i$,
${\sigma}_{ij} = I$ if the vertices $v_i$ and $v_j$ of $G$ are not
connected by an edge. Thus, in this formalism, a two-level system
is attached with each vertex of the graph and each edge of the
graph represents an interaction (ising type) between the two
two-level subsystems attached to the two vertices of the edge.

Let $G$ be a graph on $n = p.q$ vertices $v_1$, $v_2$, $\ldots$,
$v_n$. These vertices are represented here as ordered pairs in the
following way: $v_1 = (u_1, w_1) \equiv u_1w_1$, $v_2 = (u_1, w_2)
\equiv u_1w_2$, $\ldots$, $v_q = (u_1, w_q) \equiv u_1w_q$, $v_{q
+ 1} = (u_2, w_1) \equiv u_2w_1$, $v_{q + 2} = (u_2, w_2) \equiv
u_2w_2$, $\ldots$, $v_{2q} = (u_2, w_q) \equiv u_2w_q$, $\ldots
\ldots$, $v_{(p - 1)q + 1} = (u_p, w_1) \equiv u_pw_1$, $v_{(p -
1)q + 2} = (u_p, w_2) \equiv u_pw_2$, $\ldots$, $v_{pq} = (u_p,
w_q) \equiv u_pw_q$. We associate to this graph $G$ on $n$ labeled
vertices (described above) the orthonormal basis
$\{|v_{i}\rangle:i=1,2,\ldots,n\}=\{|u_{j}\rangle\otimes|w_{k}\rangle
:j=1,2,\ldots,p;k=1,2,\ldots,q\}$, where
$\{|u_{j}\rangle:j=1,2,\ldots,p\}$ and
$\{|w_{k}\rangle:k=1,2,\ldots,q\}$ are orthonormal bases of the
Hilbert
spaces $\mathcal{H}_{A}\cong\mathbb{C}^{p}$ and $\mathcal{H}_{B}%
\cong\mathbb{C}^{q}$, respectively. The \emph{partial transpose}
of a graph
$G=(V,E)$ (with respect to $\mathcal{H}_{B}$), denoted by $G^{\Gamma_{B}%
}=(V,E^{\prime})$, is the graph such that
$\{u_{i}w_{j},u_{k}w_{l}\}\in E^{\prime}$ if and only if
$\{u_{i}w_{l},u_{k}w_{j}\}\in E$. We propose the following
conjecture:

\begin{conjecture}
Let $\rho(G)$ be the density matrix of a graph on $n=pq$ vertices.
Then $\rho(G)$ is separable in
$\mathbb{C}_{A}^{p}\otimes\mathbb{C}_{B}^{q}$ if and only if
$\Delta(G)=\Delta\left(  G^{\Gamma_{B}}\right)  $.
\end{conjecture}

A proof of this conjecture would give a simple method for testing
the separability of density matrices of graphs, as we would only
need to check whether the $n\times n$ diagonal matrices
$\Delta(G)$ and $\Delta\left( G^{\Gamma_{B}}\right)  $ are equal.
This fact is in some sense analogous to the fact that the
separability of all two-mode Gaussian states (which form a
continuous family) is equivalent to the Peres--Horodecki partial
transposition criterion \cite{s}. We prove one side of our
conjecture:

\begin{theorem}
Let $\rho(G)$ be the density matrix of a graph on $n=pq$ vertices.
If $\rho(G)$ is separable in
$\mathbb{C}_{A}^{p}\otimes\mathbb{C}_{B}^{q}$ then
$\Delta(G)=\Delta\left(  G^{\Gamma_{B}}\right)  $.
\end{theorem}

We prove the other side of the conjecture for the following two
families of graphs:

\begin{itemize}
\item Consider a rectangular lattice with $pq$ points arranged in
$p$ rows and $q$ columns, such that the distance between two
neighboring points on the same row or in the same column is $1$. A
\emph{nearest point graph} is a graph whose vertices are
identified with the points of the lattice and the edges have
length $1$ or $\sqrt{2}$.

\item A \emph{perfect matching} is a graph $G=(V,E)$ such that for
every $v_{i}$ there is a unique vertex $v_{j}$ such that
$\{v_{i},v_{j}\}\in E$.
\end{itemize}

Namely, we prove the following two theorems:

\begin{theorem}
Let $G$ be a nearest point graph on $n=pq$ vertices. Then the
density matrix $\rho(G)$ is separable in
$\mathbb{C}_{A}^{p}\otimes\mathbb{C}_{B}^{q}$ iff $\Delta\left(
G\right)  =\Delta\left(  G^{\Gamma_{B}}\right)  $.
\end{theorem}

\begin{theorem}
Let $G$ be a perfect matching on $n=2k$ vertices. Then the density
matrix $\rho(G)$ is separable in $\mathbb{C}_{A}^{k}\otimes$
$\mathbb{C}_{B}^{2}$ iff $\Delta\left(  G\right)  =\Delta\left(
G^{\Gamma_{B}}\right)  $.
\end{theorem}

See Figure 1 below as examples of perfect matching $H$, the
partial transpose graph $H^{{\Gamma}_B}$, nearest point graph $G$,
and the partial transpose graph $G^{{\Gamma}_B}$.


\begin{center}
\begin{figure} [h]
  \includegraphics[scale=0.4]{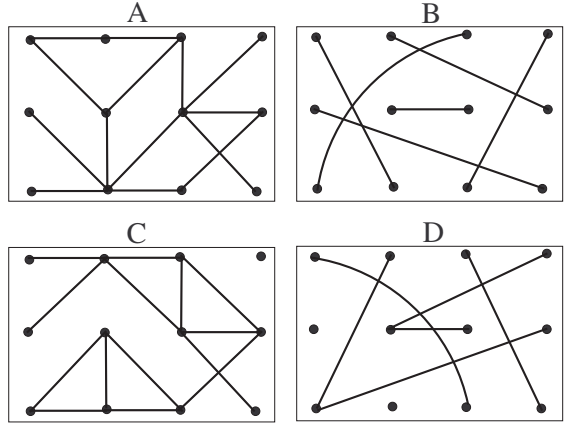}
  \caption{A) a nearest point graph $G$; B)
a perfect matching $H$; C) the partial transpose graph
$G^{\Gamma_{B}}$; D): the partial transpose graph
$H^{\Gamma_{B}}$.}\label{figure-perfect-match}
\end{figure}
\end{center}

\vspace{0.3cm} The \emph{degree condition}\ expressed in the
conjecture appears to have value beyond density matrices of
graphs. In general, given a density matrix $\rho$ in
$\mathbb{C}_{A}^{p}\otimes$ $\mathbb{C}_{B}^{q}$, let
$\Delta(\rho)$ be the matrix defined as follows:
\[
\lbrack\Delta(\rho)]_{i,j}=\left\{
\begin{tabular}
[c]{cc}%
$\sum_{k=1}^{pq}\rho_{ik},$ & if $i=j;$\\
$0,$ & if $i\neq j.$%
\end{tabular}
\right.
\]
In a \emph{circulant matrix} each row is a cyclic shift of the row
above to the right. This means that a circulant matrix is then
defined by its first row. Let $G$ be a finite group of order $n$
and let $\sigma$ be the regular permutation representation of $G$.
Then $\sigma$ is an homomorphism from $G$ to the set of
permutation matrices of dimension $n$. The \emph{fourier
transform} (evaluated at $\sigma$) of a complex-valued function
$f$ on $G$ is defined as the matrix $\widehat{f}=\sum
f(g)\sigma(g)$ \cite{terras}. According to this definition, a
complex circulant matrix $M$ of dimension $n$ has the form
$M=\sum_{g\in\mathbb{Z}_{n}}f(g)\sigma(g)$. We prove the following
result:

\begin{theorem}
Let $\rho$ be a circulant density matrix of dimension $n=pq$. Then
$\Delta(\rho)=\Delta\left(  \rho^{\Gamma_{B}}\right)  $ and $\rho$
is separable in $\mathbb{C}_{A}^{p}\otimes$ $\mathbb{C}_{B}^{q}$.
Let $\rho =\sum_{g\in\mathbb{Z}_{2}^{n}}f(g)\sigma(g)$ be a
density matrix of dimension $2^{n}$. Then
$\Delta(\rho)=\Delta\left(  \rho^{\Gamma_{B}}\right)  $ and
$\rho$ is separable in $\mathbb{C}_{A}^{2^{k}}\otimes$ $\mathbb{C}_{B}^{2^{l}%
}$, where $k+l=n$.
\end{theorem}

\section{Proofs}

\subsection{Proof of Theorem 2}

Let $L(G)$ be the laplacian of a graph $G=(V,E)$ on $n$ vertices
$v_{1},...,v_{n}$. Let $D$ be any $n\times n$ real diagonal matrix
in the orthonormal basis $\{|v_{1}\rangle,...,|v_{n}\rangle\}$
such that $D\neq0$ and tr$(D)=0$. It follows that there is at
least one negative entry in the diagonal of $D$. Let this entry be
$D_{i,i}=b_{i}$. Let $|\psi_{0}\rangle
=\sum_{j=1}^{n}|v_{j}\rangle$ and $|\phi\rangle=\sum_{j=1}^{n}\chi_{j}%
|v_{j}\rangle$, where%
\[%
\begin{tabular}
[c]{lll}%
$\chi_{j}=\left\{
\begin{array}
[c]{ll}%
0, & \text{if }j\neq i;\\
k & \text{if }j=i,
\end{array}
\right.  $ &  & with $k\in\mathbb{R}$.
\end{tabular}
\]
Let
$|\chi\rangle=|\psi_{0}\rangle+|\phi\rangle=\sum_{j=1}^{n}(1+\chi
_{j})|v_{j}\rangle$. then
\begin{align*}
\langle\chi|\left(  L(G)+D\right)  |\chi\rangle &
=\langle\chi|L(G)|\chi
\rangle+\langle\chi|D|\chi\rangle\\
&  =\langle\psi_{0}|L(G)|\psi_{0}\rangle+\langle\psi_{0}|L(G)|\phi
\rangle+\langle\phi|L(G)|\psi_{0}\rangle+\langle\phi|L(G)|\phi\rangle\\
&
+\langle\psi_{0}|D|\psi_{0}\rangle+\langle\psi_{0}|D|\phi\rangle
+\langle\phi|D|\psi_{0}\rangle+\langle\phi|D|\phi\rangle.
\end{align*}
The state $|\psi_{0}\rangle$ is an eigenvector (unnormalized) of
$L(G)$, corresponding to the eigenvalue $0$:
$L(G)|\psi_{0}\rangle=0$. Also
$\langle\psi_{0}|D|\psi_{0}\rangle=$ tr$(D)=0$. Then
\[
\langle\chi|\left(  L(G)+D\right)
|\chi\rangle=\langle\psi_{0}|L(G)|\phi
\rangle+\langle\phi|L(G)|\phi\rangle+\langle\psi_{0}|D|\phi\rangle+\langle
\phi|D|\psi_{0}\rangle+\langle\phi|D|\phi\rangle.
\]
Now $\langle\psi_{0}|L(G)|\phi\rangle=\langle\phi|L(G)^{T}|\psi_{0}%
\rangle=\langle\phi|L(G)|\psi_{0}\rangle=0$. In fact,
$L(G)=L(G)^{T} $. Let $[L(G)]_{j,l}$ be the $jl$-th entry of
$L(G)$ with respect to the basis
$\{|v_{1}\rangle,...,|v_{n}\rangle\}$. Let
$d_{i}=|\{v_{j}:\{v_{i},v_{j}\}\in
E\}|$. We have%
\begin{align*}
\langle\phi|L(G)|\phi\rangle &  =k^{2}(L(G))_{i,i}=k^{2}d_{i};\\
\langle\phi|D|\phi\rangle &  =b_{i}k^{2};\\
\langle\psi_{0}|D|\phi\rangle &  =b_{i}k;\\
\langle\phi|D|\psi_{0}\rangle &  =b_{i}k.
\end{align*}
Thus%
\[%
\begin{tabular}
[c]{lll}%
$\langle\chi|\left(  L(G)+D\right)
|\chi\rangle=k^{2}(d_{i}+b_{i})+2b_{i}k,$
&  & with $d_{i}\geq0$ and $b_{i}<0.$%
\end{tabular}
\]
So we can then always choose a positive $k$, small enough, such
that
\[
2b_{i}k+k^{2}(d_{i}+b_{i})<0.
\]
It follows that
\[
L(G)+D\ngeq0.
\]
For any graph $G$ on $n=pq$ vertices
\[
v_{1} = (u_{1}, w_{1}), v_{2} = (u_{1}, w_{2}),...,v_{q} = (u_{1},
w_{q}), v_{q+1} = (u_{2}, w_{1}), v_{q+2} = (u_{2},
w_{2}),...,v_{2q} = (u_{2}, w_{q}),...,v_{pq} = (u_{p}, w_{q}),
\]
consider now the degree condition $\Delta\left(  G\right)
=\Delta\left(
G^{\Gamma_{B}}\right)  $. Now%
\[
\left(  L(G)\right)  ^{\Gamma_{B}}=\left(  \Delta\left(  G\right)
-\Delta\left(  G^{\Gamma_{B}}\right)  \right)  +L(G^{\Gamma_{B}}).
\]
Let%
\[
D=\Delta\left(  G\right)  -\Delta\left(  G^{\Gamma_{B}}\right)  .
\]
Then $D$ is an $n\times n$ real diagonal matrix with respect to
the orthonormal basis
\[
|v_{1}\rangle=|u_{1}\rangle\otimes|w_{1}\rangle,...,|v_{pq}\rangle
=|u_{p}\rangle\otimes|w_{q}\rangle.
\]
Also
\[
\text{tr}(D)=\text{ tr}(\Delta\left(  G\right)  )-\text{
tr}(\Delta\left( G^{\Gamma_{B}}\right)  )=0.
\]

As $G^{{\Gamma}_B}$ is a graph on $n$ vertices $v_1$, $v_2$,
$\ldots$, $v_n$, as here $D = \Delta(G) - \Delta(G^{{\Gamma}_B})$
is a diagonal matrix with respect to the orthonormal basis
$\{|v_1\rangle, |v_2\rangle, \ldots, |v_n\rangle\}$, and as here
${\rm tr} (D) = 0$, therefore, by the above-mentioned reasoning,
$D + L(G^{{\Gamma}_B}) \ngeq 0$ if $D \ne 0$. Now if $\rho(G)$ is
separable then we must have ${L(G)}^{{\Gamma}_B}$ ($= D +
L(G^{{\Gamma}_B})$) $\geqslant 0$ \cite{P}. Therefore separability
of $L(G)$ implies that $D = \Delta(G) - \Delta(G^{{\Gamma}_B}) =
0$.

\subsection{Proof of Theorem 3}

Let $G$ be a nearest point graph on $n=pq$ vertices and $m$ edges.
We associate to $G$ the orthonormal basis
$\{|v_{i}\rangle:i=1,2,\ldots
,n\}=\{|u_{j}\rangle\otimes|w_{k}\rangle:j=1,2,\ldots,p;k=1,2,\ldots,q\}
$, where $\{|u_{j}\rangle:j=1,2,\ldots,p\}$ is an orthonormal
basis of $\mathbb{C}_{A}^{p}$ and
$\{|w_{k}\rangle:k=1,2,\ldots,q\}$ is an orthonormal basis of
$\mathbb{C}_{B}^{q}$. Let $j,j^{\prime}\in\{1,2,\ldots,p\}$ and
$k,k^{\prime}\in\{1,2,\ldots,q\}$. Let
${\lambda}_{jk,j^{\prime}k^{\prime}} \in\{0,1\}$ be defined as
follows:
\begin{equation}
{\lambda}_{jk,j^{\prime}k^{\prime}}=\left\{
\begin{array}
[c]{ll}%
1, & \text{if }\left\{  u_{j}w_{k},u_{j^{\prime}}w_{k^{\prime}}\right\}  \in E;\\
0, & \text{if }\left\{
u_{j}w_{k},u_{j^{\prime}}w_{k^{\prime}}\right\}  \notin E.
\end{array}
\right.  \label{eq1}%
\end{equation}
Thus, for the above-mentioned nearest point graph $G$,
${\lambda}_{jk, j^{\prime}k^{\prime}}$ can have non-zero values
only in the following cases: either (i) $j^{\prime}=j$ and
$k^{\prime}=k+1$, or (ii) $j^{\prime}=j+1$ and $k^{\prime}=k$, or
(iii) $j^{\prime}=j+1$ and $k^{\prime }=k+1$, or (iv) a
combination of some or all of the three cases (i) - (iii). Let
$\rho(G)$ and $\rho(G^{\Gamma_{B}})$ be the density matrices
corresponding
to the graphs $G$ and $G^{\Gamma_{B}}$, respectively. Thus%
\[%
\begin{tabular}
[c]{lll}%
$\rho(G)=\frac{1}{2m}\left(  \Delta(G)-M(G)\right)  $ & and &
$\rho (G^{\Gamma_{B}})=\frac{1}{2m}\left(  \Delta\left(
G^{\Gamma_{B}}\right)
-M\left(  G^{\Gamma_{B}}\right)  \right)  $%
\end{tabular}
\]
Let $G_{1}$ be the subgraph of $G$ whose edges are all the
entangled edges of $G$. An edge $\{ij,kl\}$ is \emph{entangled} if
$i\neq k$ and $j\neq l$. Also let $G_{1}^{\prime}$ be the subgraph
of $G^{\Gamma_{B}}$ corresponding to all the \textquotedblleft
entangled edges\textquotedblright\ of $G^{\Gamma_{B}}$. Obviously
$G_{1}^{\prime}=(G_{1})^{\Gamma_{B}}$. Using the above-mentioned
notations, we have%
\begin{align}
\rho(G_{1})  &  =\frac{1}{m}\sum_{j=2}^{p}\left(  {\lambda}_{(j-1)1,j2}%
P\left[  \frac{1}{\sqrt{2}}\left(  \left\vert
u_{(j-1)}w_{1}\right\rangle
-\left\vert u_{j}w_{2}\right\rangle \right)  \right]  \right) \nonumber\\
&  +\frac{1}{m}\sum_{j=2}^{p}\sum_{k=3}^{q}\left(  {\lambda}%
_{(j-1)(k-1),j(k-2)}P\left[  \frac{1}{\sqrt{2}}\left(  \left\vert
u_{j-1}w_{k-1}\right\rangle -\left\vert u_{j}w_{k-2}\right\rangle
\right)
\right]  \right. \nonumber\\
&  +\left.  {\lambda}_{(j-1)(k-1),jk}P\left[
\frac{1}{\sqrt{2}}\left( \left\vert u_{j-1}w_{k-1}\right\rangle
-\left\vert u_{j}w_{k}\right\rangle
\right)  \right]  \right) \nonumber\\
&  +\frac{1}{m}\sum_{j=2}^{p}\left(
{\lambda}_{(j-1)q,j(q-1)}P\left[ \frac{1}{\sqrt{2}}\left(
\left\vert u_{j-1}w_{q}\right\rangle -\left\vert
u_{j}w_{q-1}\right\rangle \right)  \right]  \right),  \label{eq3}%
\end{align}
where, for any normalized pure state $|\psi\rangle$,
$P[|\psi\rangle]$ denotes the one-dimensional projector onto the
vector $|\psi\rangle$.
Also we have%
\begin{align}
\rho({G}_{1}^{\prime}{)}  &  {=}\frac{1}{m}\sum_{j=2}^{p}\left(
{\lambda
}_{(j-1)1,j2}P\left[  \frac{1}{\sqrt{2}}\left(  \left\vert u_{(j-1)}%
w_{2}\right\rangle -\left\vert u_{j}w_{1}\right\rangle \right)
\right]
\right) \nonumber\\
&  +\frac{1}{m}\sum_{j=2}^{p}\sum_{k=3}^{q}\left(  {\lambda}%
_{(j-1)(k-1),j(k-2)}P\left[  \frac{1}{\sqrt{2}}\left(  \left\vert
u_{j-1}w_{k-2}\right\rangle -\left\vert u_{j}w_{k-1}\right\rangle
\right)
\right]  \right. \nonumber\\
&  +\left.  {\lambda}_{(j-1)(k-1),jk}P\left[
\frac{1}{\sqrt{2}}\left( \left\vert u_{j-1}w_{k}\right\rangle
-\left\vert u_{j}w_{k-1}\right\rangle
\right)  \right]  \right) \nonumber\\
&  +\frac{1}{m}\sum_{j=2}^{p}\left(
{\lambda}_{(j-1)q,j(q-1)}P\left[ \frac{1}{\sqrt{2}}\left(
\left\vert u_{j-1}w_{q-1}\right\rangle -\left\vert
u_{j}w_{q}\right\rangle \right)  \right]  \right)  \label{eq4}%
\end{align}
One can check that
\begin{align}
\Delta\left(  G_{1}\right)   &  =\frac{1}{2m}\left(
{\lambda}_{11,22}P\left[ \left\vert u_{1}w_{1}\right\rangle
\right]  +\sum_{k=3}^{q}\left(  {\lambda
}_{1(k-1),2(k-2)}+{\lambda}_{1(k-1),2k}\right)  P\left[
\left\vert u_{1}w_{k-1}\right\rangle \right]
+{\lambda}_{1q,2(q-1)}P\left[  \left\vert
u_{1}w_{q}\right\rangle \right]  \right) \nonumber\\
&  +\frac{1}{2m}\sum_{j=3}^{p}\left(
{\lambda}_{(j-2)2,(j-1)1}+{\lambda }_{(j-1)1,j2}\right)  P\left[
\left\vert u_{j-1}w_{1}\right\rangle \right]
\nonumber\\
&  +\frac{1}{2m}\sum_{j=3}^{p}\sum_{k=3}^{q}\left(  {\lambda}%
_{(j-2)(k-2),(j-1)(k-1)}+{\lambda}_{(j-2)k,(j-1)(k-1)}+{\lambda}%
_{(j-1)(k-1),j(k-2)}+{\lambda}_{(j-1)(k-1),jk}\right) \nonumber\\
&  \times P\left[  \left\vert u_{j-1}w_{k-1}\right\rangle \right] \nonumber\\
&  +\frac{1}{2m}\sum_{j=3}^{p}\left(
{\lambda}_{(j-2)(q-1),(j-1)q}+{\lambda }_{(j-1)q,j(q-1)}\right)
P\left[  \left\vert u_{j-1}w_{q}\right\rangle
\right] \nonumber\\
&  +\frac{1}{2m}{\lambda}_{(p-1)2,p1}P\left[  \left\vert u_{p}w_{1}%
\right\rangle \right]  +\frac{1}{2m}\sum_{k=3}^{q}\left(  {\lambda
}_{(p-1)(k-2),p(k-1)}+{\lambda}_{(p-1)k,p(k-1)}\right)  P\left[
\left\vert
u_{p}w_{k-1}\right\rangle \right] \nonumber\\
&  +\frac{1}{2m}{\lambda}_{(p-1)(q-1),pq}P\left[  \left\vert u_{p}%
w_{q}\right\rangle \right]  . \label{eq5}%
\end{align}
And%
\begin{align}
\Delta\left(  G_{1}^{\prime}\right)   &  =\frac{1}{2m}\left(
{\lambda
}_{12,21}P\left[  \left\vert u_{1}w_{1}\right\rangle \right]  +\sum_{k=3}%
^{q}\left(  {\lambda}_{1(k-2),2(k-1)}+{\lambda}_{1k,2(k-1)}\right)
P\left[ \left\vert u_{1}w_{k-1}\right\rangle \right]
+{\lambda}_{1(q-1),2q}P\left[
\left\vert u_{1}w_{q}\right\rangle \right]  \right) \nonumber\\
&  +\frac{1}{2m}\sum_{j=3}^{p}\left(
{\lambda}_{(j-2)1,(j-1)2}+{\lambda }_{(j-1)2,j1}\right)  P\left[
\left\vert u_{j-1}w_{1}\right\rangle \right]
\nonumber\\
&  +\frac{1}{2m}\sum_{j=3}^{p}\sum_{k=3}^{q}\left(  {\lambda}%
_{(j-2)(k-1),(j-1)(k-2)}+{\lambda}_{(j-2)(k-1),(j-1)k}+{\lambda}%
_{(j-1)(k-2),j(k-1)}+{\lambda}_{(j-1)k,j(k-1)}\right) \nonumber\\
&  \times P\left[  \left\vert u_{j-1}w_{k-1}\right\rangle \right] \nonumber\\
&  +\frac{1}{2m}\sum_{j=3}^{p}\left(
{\lambda}_{(j-2)q,(j-1)(q-1)}+{\lambda }_{(j-1)(q-1),jq}\right)
P\left[  \left\vert u_{j-1}w_{q}\right\rangle
\right] \nonumber\\
&  +\frac{1}{2m}{\lambda}_{(p-1)1,p2}P\left[  \left\vert u_{p}w_{1}%
\right\rangle \right]  +\frac{1}{2m}\sum_{k=3}^{q}\left(  {\lambda
}_{(p-1)(k-1),p(k-2)}+{\lambda}_{(p-1)(k-1),pk}\right)  P\left[
\left\vert
u_{p}w_{k-1}\right\rangle \right] \nonumber\\
&  +\frac{1}{2m}{\lambda}_{(p-1)q,p(q-1)}P\left[  \left\vert u_{p}%
w_{q}\right\rangle \right]  . \label{eq6}%
\end{align}
Let $G_{2}$ and $G_{2}^{\prime}$ respectively be the subgraphs of
$G$ and
$G^{\Gamma_{B}}$ each containing all the edges of the forms $\{u_{i}%
w_{j},u_{i}w_{j^{\prime}}\}$ (where $j\neq j^{\prime}$) as well as
$\{u_{i}w_{j},u_{i^{\prime}}w_{j}\}$ (where $i\neq i^{\prime}$).
Then it is obvious that $\Delta(G_{2})=\Delta(G_{2}^{\prime})$,
due to the fact that $G_{2}$ and $G_{2}^{\prime}$ represent the
same graph. So $\Delta
(G)=\Delta(G^{\Gamma_{B}})$ if and only if $\Delta(G_{1})=\Delta(G_{1}%
^{\prime})$. Using equations (\ref{eq5}) and (\ref{eq6}), we see
that the equality of $\Delta(G_{1})$ and $\Delta(G_{1}^{\prime})$
implies that:
\begin{equation}
\left.
\begin{array}
[c]{ccc}%
{\lambda}_{11,22} & = & {\lambda}_{12,21},\\
{\lambda}_{1(k-1),2(k-2)}+{\lambda}_{1(k-1),2k} & = & {\lambda}%
_{1(k-2),2(k-1)}+{\lambda}_{1k,2(k-1)},~\mathrm{for}~k=3,4,\ldots,q,\\
{\lambda}_{1q,2(q-1)} & = & {\lambda}_{1(q-1),2q};
\end{array}
\right\}  \label{eq7}%
\end{equation}%
\begin{equation}
\left.
\begin{array}
[c]{ccc}%
\mathrm{for}~\mathrm{each}~j\in\{3,4,\ldots,p\}: &  & \\
{\lambda}_{(j-2)2,(j-1)1}+{\lambda}_{(j-1)1,j2} & = & {\lambda}%
_{(j-2)1,(j-1)2}+{\lambda}_{(j-1)2,j1},\\
{\lambda}_{(j-2)(k-2),(j-1)(k-1)}+{\lambda}_{(j-2)k,(j-1)(k-1)} &
+ &
{\lambda}_{(j-1)(k-1),j(k-2)}+{\lambda}_{(j-1)(k-1),jk}\\
& = & \\
{\lambda}_{(j-2)(k-1),(j-1)(k-2)}+{\lambda}_{(j-2)(k-1),(j-1)k} &
+ &
{\lambda}_{(j-1)(k-2),j(k-1)}+{\lambda}_{(j-1)k,j(k-1)}\\
&  & \mathrm{for}~k=3,4,\ldots,q,\\
{\lambda}_{(j-2)(q-1),(j-1)q}+{\lambda}_{(j-1)q,j(q-1)} & = &
{\lambda }_{(j-2)q,(j-1)(q-1)}+{\lambda}_{(j-1)(q-1),jq};
\end{array}
\right\}  \label{eq8}%
\end{equation}%
\begin{equation}
\left.
\begin{array}
[c]{ccc}%
{\lambda}_{(p-1)1,p2} & = & {\lambda}_{(p-1)2,p1},\\
{\lambda}_{(p-1)(k-2),p(k-1)}+{\lambda}_{(p-1)k,p(k-1)} & = &
{\lambda
}_{(p-1)(k-1),p(k-2)}+{\lambda}_{(p-1)(k-1),pk},\\
&  & \mathrm{for}~k=3,4,\ldots,q,\\
{\lambda}_{(p-1)(q-1),pq} & = & {\lambda}_{(p-1)q,p(q-1)}.
\end{array}
\right\}  \label{eq9}%
\end{equation}
The solution of equations (\ref{eq7}) - (\ref{eq9}) is of the
form:
\begin{equation}%
\begin{tabular}
[c]{lll}%
${\lambda}_{ij,i^{\prime}j^{\prime}}={\lambda}_{ij^{\prime},i^{\prime}j},$
&
& for all $i,i^{\prime}\in\{1,2,\ldots,p\}$ and all $j,j^{\prime}%
\in\{1,2,\ldots,q\}$,
\end{tabular}
\label{eq10}%
\end{equation}
and where ever ${\lambda}_{ij,i^{\prime}j^{\prime}}$ and ${\lambda
}_{ij^{\prime},i^{\prime}j}$ are defined. Equation (\ref{eq10})
shows that
whenever there is an entangled edge $\{u_{i}w_{j},u_{i^{\prime}}w_{j^{\prime}%
}\}$ in $G$ (so we must have $i\neq i^{\prime}$ and $j\neq
j^{\prime}$), there must be the entangled edge
$\{u_{i}w_{j^{\prime}},u_{i^{\prime}}w_{j}\}$ in $G$. The two
entangled edges $\{u_{i}w_{j},u_{i^{\prime}}w_{j^{\prime}}\}$ and
$\{u_{i}w_{j^{\prime}},u_{i^{\prime}}w_{j}\}$ in $G$ together give
rise to the following contribution (which is again a density
matrix) in the density matrix ${\rho}(G)$, with the multiplicative
factor $\frac{2}{m}$:
\begin{equation}
\rho\left(  i,i^{\prime};j,j^{\prime}\right)  =\frac{1}{2}\left(
P\left[ \frac{1}{\sqrt{2}}\left(  \left\vert
u_{i}w_{j}\right\rangle -\left\vert
u_{i^{\prime}}w_{j^{\prime}}\right\rangle \right)  \right]
+P\left[  \frac {1}{\sqrt{2}}\left(  \left\vert
u_{i}w_{j^{\prime}}\right\rangle -\left\vert
u_{i^{\prime}}w_{j}\right\rangle \right)  \right]  \right)  . \label{eq11}%
\end{equation}
Let us write
\begin{equation}%
\begin{tabular}
[c]{lll}%
$\frac{1}{\sqrt{2}}\left(  \left\vert u_{i}\right\rangle
\pm\left\vert
u_{i^{\prime}}\right\rangle \right)  =\left\vert V\left(  i,i^{\prime}%
;\pm\right)  \right\rangle $ & and & $\frac{1}{\sqrt{2}}\left(
\left\vert w_{j}\right\rangle \pm\left\vert
w_{j^{\prime}}\right\rangle \right)
\left\vert X\left(  j,j^{\prime};\pm\right)  \right\rangle .$%
\end{tabular}
\label{eq12}%
\end{equation}
Using Equation (\ref{eq12}), it is easy to see from Equation
(\ref{eq11}) that
\begin{equation}
\rho\left(  i,i^{\prime};j,j^{\prime}\right)  =\frac{1}{2}P\left[
\left\vert V\left(  i,i^{\prime};+\right)  X\left(
j,j^{\prime};-\right)  \right\rangle \right]  +\frac{1}{2}P\left[
\left\vert V\left(  i,i^{\prime};-\right)
X\left(  j,j^{\prime};+\right)  \right\rangle \right]  , \label{eq13}%
\end{equation}
which is a separable state in $\mathbb{C}_{A}^{p}\otimes$ $\mathbb{C}_{B}^{q}%
$. This shows that, under the constraint
$\Delta(G_{1})=\Delta(G_{1}^{\prime })$, ${\rho}(G_{1})$ is
nothing but equal mixture of separable states of the form
$\rho(i,i^{\prime};j,j^{\prime})$, and so, ${\rho}(G_{1})$ must be
separable, which, in turn, shows that ${\rho}(G)$ has to be
separable. {This
shows that a nearest point graph }$G$ is separable in $\mathbb{C}_{A}%
^{p}\otimes$ $\mathbb{C}_{B}^{q}$ if and only if $\Delta(G)=\Delta
(G^{\Gamma_{B}})$. - We invite the reader to give a shorter proof!
-

\subsection{Perfect matchings}

{\noindent {\bf (C.1) Proof of Theorem 4}}

\vspace{0.4cm}

{\noindent {\textbf {Definition (degree condition):}}} {\it For
any graph $G$ on $n = pq$ vertices $v_1 \equiv (u_1, w_1)$, $v_2
\equiv (u_1, w_2)$, $\ldots$, $v_q \equiv (u_1, w_q)$, $v_{(q +
1)} \equiv (u_2, w_1)$, $v_{(q + 2)} \equiv (u_2, w_2)$, $\ldots$,
$v_{2q} \equiv (u_2, w_q)$, $\ldots$, $\ldots$, $v_{(p - 1)q + 1}
\equiv (u_p, w_1)$, $v_{(p - 1)q + 2} \equiv (u_p, w_2)$,
$\ldots$, $v_{pq} \equiv (u_p, w_q)$, the equation $\Delta(G) =
\Delta\left(G^{{\Gamma}_B}\right)$ is called as the degree
condition, where $G^{{\Gamma}_B}$ is the graph with
$V\left(G^{{\Gamma}_B}\right) = V(G)$ and $\{(u_i, w_j),
(u_{i^{\prime}}, w_{j^{\prime}})\} \in
E\left(G^{{\Gamma}_B}\right)$ if and only if $\{(u_i,
w_{j^{\prime}}), (u_{i^{\prime}}, w_j)\} \in E(G)$.}

\vspace{0.3cm}

We consider here only those graphs $G$ on $n=pq$ vertices, where
$n$ is even and $E(G)$ consists of edges of the forms $\{(i_{k},
j_{k}), (i_{k}^{\prime}, j_{k}^{\prime})\}$, where (i) $k$ runs
from $1$ up to $n/2$, (ii) $(i_k, j_k) \ne (i_k^{\prime},
j_k^{\prime})$ for all $k$, (iii) $(i_k, j_k) \ne (i_l, j_l)$
whenever $k \ne l$, (iv) $(i_k, j_k) \ne (i_l^{\prime},
j_l^{\prime})$ whenever $k \ne l$, and (v) $(i_k^{\prime},
j_k^{\prime}) \ne (i_l^{\prime}, j_l^{\prime})$ whenever $k \ne
l$. Thus $G$ is nothing but a perfect matching on $n = pq$
vertices. In addition to above-mentioned conditions, if we have
$i_k \ne i_k^{\prime}$ and $j_k \ne j_k^{\prime}$ for each $k \in
\{1, 2, \ldots, n/2\}$, then $G$ is called as a {\it perfect
entangling matching}. We denote the set of all such perfect
entangling matchings on the same set of $n=pq$ vertices as
$\mathcal{P}_{p.q}$. The density matrix ${\rho}(G)$ of the graph
$G$ is given by
\[
\rho(G)=\frac{2}{n}\sum_{k=1}^{n/2}P\left[
\frac{1}{\sqrt{2}}\left( \left\vert i_{k}j_{k}\right\rangle
-\left\vert i_{k}^{\prime}j_{k}^{\prime }\right\rangle \right)
\right]  .
\]

Let $G\in\mathcal{P}_{p.q}$. Let $G^{\Gamma_{B}}$ be the graph
with vertex set as $V(G)$ and $\{(i_{k}, j_{k}^{\prime}),
(i_{k}^{\prime}, j_{k})\}\in
E(G^{\Gamma_{B}%
})$ if and only if $\{(i_{k}, j_{k}), (i_{k}^{\prime},
j_{k}^{\prime})\}\in E(G)$. Let
$\mathcal{P}^S_{p.q}=\{G\in\mathcal{P}_{p.q}:G^{\Gamma_{B}}\in\mathcal{P}%
_{p.q}\}$. It can be easily shown that for any perfect matching $G$
on $n = p.q$ vertices, $G \in {\cal P}^S_{p.q}$ if and only if
$\Delta(G) = \Delta\left(G^{{\Gamma}_B}\right)$. Following are the
two examples of `canonical' perfect entangling matchings:

\vspace{0.2cm} {\noindent {\bf (1) Cris-cross:} A cris-cross ${\cal
C}$ is given by ${\cal C} = \left(V({\cal C}) = \left\{\left(i_1,
1\right), \left(i_1, 2\right), \left(i_2, 1\right), \left(i_2,
2\right)\right\}, E({\cal C}) = \left\{\left\{\left(i_1, 1\right),
\left(i_2, 2\right)\right\}, \left\{\left(i_2, 1\right), \left(i_1,
2\right)\right\}\right\}\right)$ (where $i_1 \ne i_2$).}

{\noindent {\bf (2) Tally mark:} A tally mark ${\cal T}$ is given by
${\cal T} = (V({\cal T}), E({\cal T}))$ where $V({\cal T}) =
\left\{\left(i_k, 1\right) : k = 1, 2, \ldots, r^{\prime}\right\}
\bigcup \left\{\left(i_k, 2\right) : k = 1, 2, \ldots,
r^{\prime}\right\}$, and $E({\cal T}) = \left\{\left\{\left(i_1,
1\right), \left(i_2, 2\right)\right\}, \left\{\left(i_2, 1\right),
\left(i_3, 2\right)\right\}, \ldots, \left\{\left(i_{(r^{\prime} -
1)}, 1\right), \left(i_{r^{\prime}}, 2\right)\right\},
\left\{\left(i_{r^{\prime}}, 1\right), \left(i_1,
2\right)\right\}\right\}$ (where $1 \le i_1 < i_2 < \ldots <
i_{r^{\prime}} \le p^{\prime}$ and $r^{\prime} \le p^{\prime}$).}

We are now ready to give a proof of Theorem 4.

\vspace{0.3cm}

{\noindent {\bf {Proof of Theorem 4 :}}} Let $G$ be a perfect
matching on $n = 2p$ vertices $v_1 \equiv (1, 1)$, $v_2 \equiv (1,
2)$, $v_3 \equiv (2, 1)$, $v_4 \equiv (2, 2)$, $\ldots$, $v_{(2p -
1)} \equiv (p, 1)$, $v_{2p} \equiv (p, 2)$.

Let us first assume that $\rho(G)$ is separable in
${\mathbb{C}}^p_A \otimes {\mathbb{C}}^2_B$. Then by Theorem 2, we
have $\Delta(G) = \Delta\left(G^{{\Gamma}_B}\right)$.

Next we assume that $\Delta(G) =
\Delta\left(G^{{\Gamma}_B}\right)$. Let us denote the subgraph of
$G$, consisting of all its unentangled edges, as $G_1$ and the
subgraph of $G$, consisting of all its entangled edges, as $G_2$.
As $G$ is a perfect matching, therefore $G$ is the disjoint union
of $G_1$ and $G_2$: $G = G_1 \biguplus G_2$. Thus $V(G)$ is the
set wise disjoint union of $V\left(G_1\right)$ and
$V\left(G_2\right)$, while $E(G)$ is the set wise disjoint union
of $E\left(G_1\right)$ and $E\left(G_2\right)$. Let us take
$E\left(G_2\right) = \left\{\left\{\left(i_k, 1\right), \left(j_k,
2\right)\right\} : k = 1, 2, \ldots, q\right\}$, where $q$ is a
non-negative integer with $q \le p$, $1 \le i_1 < i_2 < \ldots <
i_q \le p$, $1 \le j_1, j_2, \ldots, j_q \le p$, $i_k \ne j_k$
whenever $k = 1, 2, \ldots, q$, and $j_k \ne j_l$ whenever $k \ne
l$. Thus we see that $V\left(G_2\right)$ is the (disjoint) union
of $\left\{\left(i_k, 1\right) : k = 1, 2, \ldots, q\right\}$ and
$\left\{\left(j_k, 2\right) : k = 1, 2, \ldots, q\right\}$.

Let us denote the subgraph of $G^{{\Gamma}_B}$, consisting of all
its unentangled edges, as $G_3$, while the subgraph of
$G^{{\Gamma}_B}$, consisting of all its entangled edges, is
denoted here by $G_4$. Here $G_1^{{\Gamma}_B} = G_1 = G_3$. This
is true for any general graph $G$ on $n = p^{\prime}.q^{\prime}$
vertices. Again, for any perfect matching $G$ on $n =
p^{\prime}.q^{\prime}$ vertices $v_1^{\prime} \equiv \left(u_1,
w_1\right)$, $v_2^{\prime} \equiv \left(u_1, w_2\right)$,
$\ldots$, $v_{q^{\prime}}^{\prime} \equiv \left(u_1,
w_{q^{\prime}}\right)$, $v_{(q^{\prime} + 1)}^{\prime} \equiv
\left(u_2, w_1\right)$, $v_{(q^{\prime} + 2)}^{\prime} \equiv
\left(u_2, w_2\right)$, $\ldots$, $v_{2q^{\prime}}^{\prime} \equiv
\left(u_2, w_{q^{\prime}}\right)$, $\ldots$, $\ldots$,
$v_{((p^{\prime} - 1)q^{\prime} + 1)}^{\prime} \equiv
\left(u_{p^{\prime}}, w_1\right)$, $v_{((p^{\prime} - 1)q^{\prime}
+ 2)}^{\prime} \equiv \left(u_{p^{\prime}}, w_2\right)$, $\ldots$,
$v_{p^{\prime}q^{\prime}}^{\prime} \equiv \left(u_{p^{\prime}},
w_{q^{\prime}}\right)$, the degree condition $\Delta(G) =
\Delta\left(G^{{\Gamma}_B}\right)$ implies that (and, is implied
by) $G^{{\Gamma}_B}$ is a perfect matching on the above-mentioned
$n = p^{\prime}.q^{\prime}$ vertices. So, we must have,
$G^{{\Gamma}_B} = G_3 \biguplus G_4$, and hence,
$V\left(G_4\right) = V\left(G_2\right)$ (this is true for
arbitrary values of $p^{\prime}$ and $q^{\prime}$ provided $n =
p^{\prime}.q^{\prime}$ is even). Thus we see that both $G_2$ as
well as $G_4$ are perfect entangling matchings on the {\it same}
subset of vertices of $G$ (this is also true for arbitrary values
of $p^{\prime}$ and $q^{\prime}$ provided $n =
p^{\prime}.q^{\prime})$. It then follows that the two subsets
$\left\{i_k : k = 1, 2, \ldots, q\right\}$ and $\left\{j_k : k =
1, 2, \ldots, q\right\}$ of $\{1, 2, \ldots, p\}$ must be same.
This is so because if some $j_k \notin \left\{i_k : k = 1, 2,
\ldots, q\right\}$, then vertex $\left(j_k, 1\right)$ of the
(entangled) edge $\left\{\left(i_k, 2\right), \left(j_k,
1\right)\right\}$ in $G_4$ will belong to $V\left(G_1\right)$ (and
hence, to $V\left(G_3\right)$) -- a contradiction. Therefore,
$G_2$ (and hence, $G_4$) is a perfect entangling matching on the
set of $2q$ vertices $(i, j)$, where $i \in \left\{i_1, i_2,
\ldots, i_q\right\}$ and $j \in \{1, 2\}$. Note that this fact is
true not only for $n = 2.p$ but for any general $n =
p^{\prime}.q^{\prime}$, provided $n$ is even (and so, for any $G
\in {\cal P}_{p^{\prime}.q^{\prime}}$, $G \in {\cal
P}^S_{p^{\prime}.q^{\prime}}$ if and only if $\Delta(G) =
\Delta\left(G^{{\Gamma}_B}\right)$).

Now, it is known that (see Lemma 4.4 in \cite{b}) any perfect
entangling matching $G^{\prime}$ on $n = 2p^{\prime}$ vertices
$v_1^{\prime} \equiv (1, 1)$, $v_2^{\prime} \equiv (1, 2)$,
$v_3^{\prime} \equiv (2, 1)$, $v_4^{\prime} \equiv (2, 2)$,
$\ldots$, $v_{(2p^{\prime} - 1)}^{\prime} \equiv (p^{\prime}, 1)$,
$v_{2p^{\prime}}^{\prime} \equiv (p^{\prime}, 2)$ can be transformed
in to a `canonical' perfect entangling matching $G_0$ on the same
set of vertices by applying a suitable permutation on the first
label of the vertices $v_1^{\prime}$, $v_2^{\prime}$, $\ldots$,
$v_{2p^{\prime}}$, where, by `canonical' perfect entangling
matching, we mean either (i) a cris-cross, or (ii) a tally mark, or
(iii) a disjoint union of some tally-marks and/or some cris-crosses
(this kind of result is still lacking for a general $G \in {\cal
P}^S_{p^{\prime}.q^{\prime}}$ and we don't know what should be the
`canonical' form of such a $G$). As $\rho\left(G_0\right)$ is known
to be separable in ${\mathbb{C}}^{p^{\prime}}_A \otimes
{\mathbb{C}}^2_B$ (according to Lemma 4.5 in \cite{b} together with
the fact that the density matrix of a cris-cross is always
separable), therefore $\rho(G^{\prime})$ is separable in
${\mathbb{C}}^{p^{\prime}}_A \otimes {\mathbb{C}}^2_B$.

Thus, it follows that $\rho\left(G_2\right)$ is separable in
${\mathbb{C}}^q_A \otimes {\mathbb{C}}^2_B$ (and hence, in
${\mathbb{C}}^p_A \otimes {\mathbb{C}}^2_B$, as the orthonormal
basis $\left\{\left|i_k\right\rangle : k = 1, 2, \ldots,
q\right\}$ of ${\mathbb{C}}^q_A$ is contained inside the
orthonormal basis $\{|l\rangle : l = 1, 2, \ldots, p\}$ of
${\mathbb{C}}^p_A$). Also $\rho\left(G_1\right)$ is separable in
${\mathbb{C}}^p_A \otimes {\mathbb{C}}^2_B$, as $G_1$ consists of
only unentangled edges of $G$. Now $\rho(G) = \frac{1}{p}\left[(p
- q)\rho\left(G_1\right) + q\rho\left(G_2\right)\right]$. Hence
$\rho(G)$ is separable in ${\mathbb{C}}^p_A \otimes
{\mathbb{C}}^2_B$.~ $\Box$

\vspace{0.2cm} During the proof of Theorem 4, we have proved the
following result:

\vspace{0.2cm} {\noindent {\bf Corollary 1:} {\it Let $G$ be a
perfect matching on $n = p.q$ vertices $v_1 = (u_1, w_1)$, $v_2 =
(u_1, w_2)$, $\ldots$, $v_n = (u_p, w_q)$ for which $\Delta(G) =
\Delta(G^{{\Gamma}_B})$. Then $G$ is a disjoint union of $N$ no.
of perfect matchings $G_1$, $G_2$, $\ldots$, $G_N$, where (i)
$V(G_i) = \{(u_{ij}, w_{ik}): j = 1, 2, \ldots, p_i; k = 1, 2,
\ldots, q_i\}$, (ii) $\bigcup_{i = 1}^{N} \{u_{ij} | j = 1, 2,
\ldots, p_i\} = \{u_1, u_2, \ldots, u_p\}$ and $\bigcup_{i =
1}^{N} \{w_{ik} | k = 1, 2, \ldots, q_i\} = \{w_1, w_2, \ldots,
w_q\}$, (iii) for any two different $i, i^{\prime} \in \{1, 2,
\ldots, N\}$, either $\{u_{ij} | j = 1, 2, \ldots, p_i\} \bigcap
\{u_{i^{\prime}j} | j = 1, 2, \ldots, p_{i^{\prime}}\} =
\varnothing$ or $\{w_{ik} | k = 1, 2, \ldots, q_i\} \bigcap
\{w_{i^{\prime}k} | k = 1, 2, \ldots, q_{i^{\prime}}\} =
\varnothing$ or both, and (iv) for each $i \in \{1, 2, \ldots,
N\}$, either $G_i$ consists of only entangled edges or only
unentangled edges, but not both.}

\vspace{0.2cm} Thus we see that, for any perfect matching $G$,
when the degree condition is satisfied, it is enough to study the
separability of the density matrices of its pairwise disjoint
entangled subgraphs ({\it i.e.}, subgraphs each of whose edge is
entangled), each of which is a perfect entangling matching on its
own right ({\it i.e.}, it is a perfect entangling matching on a
set $S$ of vertices taken from $V(G)$ such that all the elements
of $S$ can be labelled by two labels). See Figure 2 for an
illustration.

\vspace{0.3cm}
\begin{center}
\begin{figure} [h]
  \includegraphics[scale=0.4]{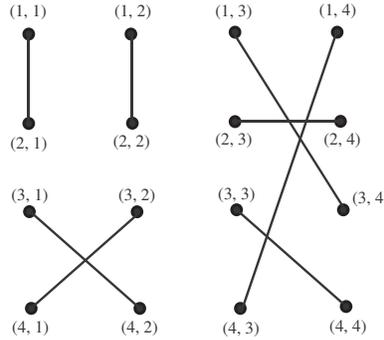}
  \caption{A perfect matching $G$ on $16$ vertices $(1, 1)$, $(1,
  2)$, $\ldots$ $(4,4)$, for which the degree condition is satisfied.
  $G$ is the disjoint union of of the following four perfect matchings:
  (i) the unentangled graph: $G_1 = (V(G_1), E(G_1))$ with $V(G_1) = \{(1,
  1), (1, 2), (2, 1), (2, 2)\}$, and $E(G_1) = \{\{(1, 1), (2, 1)\},
  \{(1, 2), (2, 2)\}\}$, (ii) the cris-cross: $G_2 = (V(G_2),
  E(G_2))$ with $V(G_2) = \{(3, 1), (3, 2), (4, 1), (4, 2)\}$ and
  $E(G_2) = \{\{(3, 1), (4, 2)\}, \{(3, 2), (4, 1)\}\}$, (iii) the
  perfect entangling matching: $G_3 = (V(G_3), E(G_3))$ with $V(G_3)
  = \{(1, 3), (1, 4), (3, 3), (3, 4), (4, 3), (4, 4)\}$ and $E(G_3)
  = \{\{(1, 3), (3, 4)\}, \{(1, 4), (4, 3)\}, \{(3, 3), (4, 4)\}\}$,
  and (iv) the unentangled graph: $G_4 = (V(G_4), E(G_4))$ with
  $V(G_4) = \{(2, 3), (2, 4)\}$ and $E(G_4) = \{(2, 3), (2,
  4)\}$.} \label{figure:all-cases}
\end{figure}

\end{center}

\vspace{0.3cm} A speciality of the case $n = p.2$ is also reflected
in the following Lemma.

\vspace{0.2cm}
{\noindent {\bf {Lemma 1:}}} ${\cal P}_{p.2} =
{\cal P}^S_{p.2}$.

{\noindent {\bf {Proof:}}} By definition, ${\cal P}^S_{p.2}
\subseteqq {\cal P}_{p.2}$. Let $G \in {\cal P}_{p.2}$, where
$V(G) = \{(k, 1) : k = 1, 2, \ldots, p\} \bigcup \{(k, 2) : k = 1,
2, \ldots, p\}$ and $E(G) = \{\{(k, 1), (i_k, 2)\} : k = 1, 2,
\ldots, p\}$ where (i) for each $k \in \{1, 2, \ldots, p\}$, $i_k$
is a particular element in $\{1, 2, \ldots, p\} \backslash \{k\}$
and (ii) $i_k \ne i_l$ whenever $k \ne l$. Thus we see that
$G^{{\Gamma}_B}$ is a graph on $2p$ vertices such that
$V\left(G^{{\Gamma}_B}\right) = V(G)$ and
$E\left(G^{{\Gamma}_B}\right) = \{\{(k, 2), (i_k, 1)\} : \{(k, 1),
(i_k, 2)\} \in E(G)~ {\rm for}~ {\rm all}~ k = 1, 2, \ldots, p\}$
with the properties that (i) for each $k \in \{1, 2, \ldots, p\}$,
$i_k$ is a particular element in $\{1, 2, \ldots, p\} \backslash
\{k\}$, (ii) $i_k \ne i_l$ whenever $k \ne l$. So $G^{{\Gamma}_B}$
must be a perfect entangling matching with vertex set as $V(G)$.
Therefore, ${\cal P}_{p.2} \subseteqq {\cal P}^S_{p.2}$.~ $\Box$

\vspace{0.2cm} The result in Lemma 1 can not, in general, be
extended for the case of ${\cal P}_{p.q}$ if $q > 2$ (see, for
example, figures 2 and 3 in \cite{b}).

\vspace{0.4cm}

{\noindent {\bf (C.2) Properties of general perfect entangling
matchings}}

\vspace{0.3cm}

If $\rho(H)$ is separable in ${\mathbb{C}}^p_A \otimes
{\mathbb{C}}^q_B$, where $H$ is the subgraph of a perfect
entangling matching $G$ on $pq$ vertices such that $H$ consists of
all the entangled edges in $G$, then $\rho(G)$ will be
automatically separable in ${\mathbb{C}}^p_A \otimes
{\mathbb{C}}^q_B$. So the relevant question is: what can we say
about separability of $\rho(G)$ whenever $G \in {\cal P}^S_{p.q}$,
with $q > 2$? Note that it is irrelevant to consider separability
of $\rho(G)$ for an arbitrary $G \in {\cal P}_{p.q}$, as $\rho(G)$
is inseparable if $G \in {\cal P}_{p.q} \backslash {\cal
P}_{p.q}^S$ (because, in that case, the degree condition is not
satisfied). As we have mentioned during the proof of Theorem 4, we
still don't have a `canonical' set of perfect entangling matchings
on $n = p.q$ vertices, to one (or a disjoint mixture of some) of
which, any element of ${\cal P}_{p.q}$ can be transformed via
local permutation(s) on one or both the labels the vertices.
Moreover, even if we have that canonical set, we still don't have
any proof of separability of the corresponding density matrices.
But for a particular class of perfect entangling matchings $G$ on
$n = p.(2r)$ vertices, for each of which the degree condition is
satisfied, one can show that $\rho(G)$ is separable in
${\mathbb{C}}^p_A \otimes {\mathbb{C}}^{2r}_B$:

Let $G \in {\cal P}_{p.(2r)}$, where $G = \biguplus_{k = 1}^{r}
G_{j_k l_k}$, with $V\left(G_{j_k l_k}\right) = \{(a, j_k) : a =
1, 2, \ldots, p\} \bigcup \{(a, l_k): a = 1, 2, \ldots, p\}$ and
$E\left(G_{j_k l_k}\right) = \{\{(a, j_k), (i_a^{(k)}, l_k)\} : a
= 1, 2, \ldots, p\}$ such that for each $k \in \{1, 2, \ldots,
r\}$, (i) $i_a^{(k)} \in \{1, 2, \ldots, p\} \backslash \{a\}$ for
each $a \in \{1, 2, \ldots, p\}$ and (ii) $j_k, l_k \in \{1, 2,
\ldots, 2r\}$ with the properties that $j_k \ne l_k$, $j_k \ne
j_{k^{\prime}}$ (if $k \ne k^{\prime}$), $l_k \ne l_{k^{\prime}}$
(if $k \ne k^{\prime}$). Thus we see that for each $k \in \{1, 2,
\ldots, r\}$, $G_{j_k l_k}$ is a perfect entangling matching on
$2p$ vertices $(1, j_k)$, $(1, l_k)$, $(2, j_k)$, $(2, l_k)$,
$\ldots$, $(p, j_k)$, $(p, l_k)$ and with $p$ edges $\{(1, j_k),
(i^{(k)}_1, l_k)\}$, $\{(2, j_k), (i^{(k)}_2, l_k)\}$, $\ldots$,
$\{(p, j_k), (i^{(k)}_p, l_k)\}$. So, by Lemma 4.4 of \cite{b},
$G_{j_k l_k}$ can be transformed into a canonical perfect
entangling matching on same set $V\left(G_{j_k l_k}\right)$ of
vertices. And so, $\rho\left(G_{j_k l_k}\right)$ is separable in
${\mathbb{C}}^p_A \otimes {\mathbb{C}}^2_B$ (and so, by Theorem 2,
$\Delta\left(G_{j_k l_k}\right) = \Delta\left(G_{j_k
l_k}^{{\Gamma}_B}\right)$). Therefore, $\rho(G)$ ($= \bigoplus_{k
= 1}^{r} \rho\left(G_{j_k l_k}\right)$) is separable in
${\mathbb{C}}^p_A \otimes {\mathbb{C}}^{(2r)}_B$ and $\Delta(G) =
\bigoplus_{k = 1}^{r} \Delta\left(G_{j_k l_k}\right) =
\bigoplus_{k = 1}^{r} \Delta\left(G_{j_k l_k}^{{\Gamma}_B}\right)
= \Delta\left(G^{{\Gamma}_B}\right)$.

The set of all elements in ${\cal P}^S_{p.(2r)}$, each of which is
a disjoint union of exactly $r$ number of elements of ${\cal
P}_{p.2}$, is denoted here by ${\cal E}_{p.(2r)}$. Let $G \in
{\cal E}_{p.(2r)}$. Then, as described above, $G$ is a disjoint
union of $r$ elements $G_{j_1 l_1}$, $G_{j_2 l_2}$, $\ldots$,
$G_{j_r l_r}$ of ${\cal P}^S_{p.2}$. Note that each element
$G_{j_k, l_k}$ of ${\cal P}^S_{p.2}$ is a disjoint union of $N_k$
number of elements $G_{1j_kl_k}\left(L_{1k}\right)$ ($\in {\cal
P}_{L_{1k}.2}$), $G_{2j_kl_k}\left(L_{2k}\right)$ ($\in {\cal
P}_{L_{2k}.2}$), $\ldots$, $G_{N_kj_kl_k}\left(L_{N_kk}\right)$
($\in {\cal P}_{L_{N_kk}.2}$) such that {\it no} further splitting
of any $G\left(L_{ik}\right)$ (as a disjoint union of perfect
entangling matchings) is possible (see Figure 3 for an
illustration). For each $i \in \{1, 2, \ldots, N_k\}$, we must
have $V\left(G_{ij_kl_k}\left(L_{ik}\right)\right) =
\left\{\left(a^{(ik)}_m, j_k\right) : m = 1, 2, \ldots,
L_{ik}\right\}\bigcup \left\{\left(a^{(ik)}_m, l_k\right) : m = 1,
2, \ldots, L_{ik}\right\}$, where $\left\{a^{(ik)}_m : m = 1, 2,
\ldots, L_{ik}\right\} \bigcap \left\{a^{(i^{\prime}k)}_m : m = 1,
2, \ldots, L_{i^{\prime}k}\right\} = \varnothing$ if $i \ne
i^{\prime}$ and $\bigcup_{i = 1}^{N_k} \left\{a^{(ik)}_m : m = 1,
2, \ldots, L_{ik}\right\} = \{1, 2, \ldots, p\}$. Now by using
Lemma 4.4 of \cite{b}, we have
$$\rho\left(G_{i^{\prime}j_kl_k}\left(L_{i^{\prime}k}\right)\right) =
\frac{1}{L_{i^{\prime}k}} \sum_{l = 0}^{L_{i^{\prime}k} - 1}
P\left[\frac{1}{\sqrt{L_{i^{\prime}k}}} \sum_{m =
1}^{L_{i^{\prime}k}}~ {\rm exp}~ \left(\frac{2 \pi i (m - 1)
l}{L_{i^{\prime}k}}\right)
U_{i^{\prime}k}\left|a^{(i^{\prime}k)}_m\right\rangle\right]$$
\begin{equation}
\label{eqn1} \otimes P\left[\frac{1}{\sqrt{2}}
\left(\left|j_k\right\rangle -~ {\rm exp}~ \left(-\frac{2 \pi i
l}{L_{i^{\prime}k}}\right) \left|l_k\right\rangle\right)\right],
\end{equation}
where $U_{i^{\prime}k}$ is the permutation matrix corresponding to
a permutation on the labels $a^{(i^{\prime}k)}_1$,
$a^{(i^{\prime}k)}_2$, $\ldots$,
$a^{(i^{\prime}k)}_{L_{i^{\prime}k}}$ for $i^{\prime} \in \{1, 2,
\ldots, N_k\}$ and $k = 1, 2, \ldots, r$. So, we have
$$\rho\left(G_{j_kl_k}\right) = \frac{1}{N_k}\sum_{i^{\prime} =
1}^{N_k}
\rho\left(G_{i^{\prime}j_kl_k}\left(L_{i^{\prime}k}\right)\right),$$
and finally
\begin{equation}
\label{eqn2} \rho(G) = \frac{1}{r}\sum_{k = 1}^{r}
\rho\left(G_{j_kl_k}\right).
\end{equation}

\vspace{0.3cm}
\begin{center}
\begin{figure}[h]
  \includegraphics[scale=0.4]{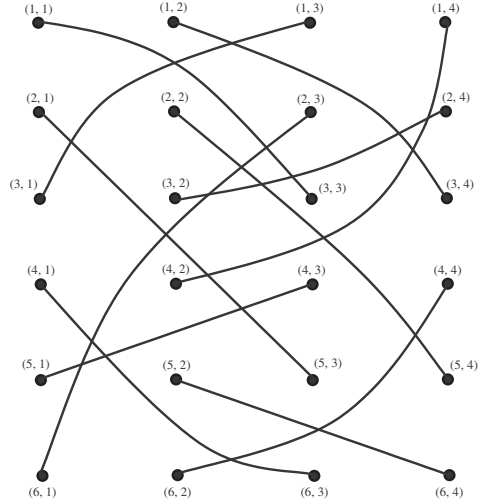}\\
  \caption{A perfect
entangling matching $G \in {\cal P}_{6.(2.2)}^S$ on the $24$
vertices $(1, 1)$, $(1, 2)$, $\ldots$, $(6, 4)$. $G$ is a disjoint
union of two perfect entangling matchings $G_{13}$ and $G_{24}$ in
${\cal P}_{6.2}$, where $V(G_{13}) = \{(i, 1) | i = 1, 2, \ldots,
6\} \bigcup \{(i, 3) | i = 1, 2, \ldots, 6\}$ and $V(G_{24}) =
\{(i, 2) | i = 1, 2, \ldots, 6\} \bigcup \{(i, 4) | i = 1, 2,
\ldots, 6\}$. Thus $G \in {\cal E}_{6.(2.2)}$. $G_{13}$ itself is
a disjoint union of the cris-cross $G_{113}(2)$ (with
$V(G_{113}(2)) = \{(1, 1), (3, 1)\} \bigcup \{(1, 3), (3, 3)\}$
and $E(G(2)) = \{\{(1, 1), (3, 3)\}, \{(3, 1), (1, 3)\}\}$) and
the perfect entangling matching $G_{213}(4)$ (with $V(G_{213}(4))
= \{(2, 1), (4, 1), (5, 1), (6, 1)\} \bigcup \{(2, 3), (4, 3), (5,
3), (6, 3)\}$ and $E(G_{213}(4)) = \{\{(2, 1), (5, 3)\}, \{(4, 1),
(6, 3)\}, \{(5, 1), (4, 3)\}, \{(6, 1), (2, 3)\}\}$). $G_{213}(4)$
can be transformed (via the local permutation $4 \leftrightarrow
5$ on the first label) to a tally mark. And $G_{24}$ is a perfect
entangling matching on the vertices $(1, 2)$, (2, 2), $\ldots$,
$(6, 2)$, $(1, 4)$, $(2, 4)$, $\ldots$, $(6, 4)$ which can be
transformed (via first applying the local permutation $2
\leftrightarrow 3, 4 \leftrightarrow 5$ and then applying the
local permutation $5 \leftrightarrow 6$, both on the first label)
to a tally mark.} \label{figure-p-e-matching}
\end{figure}
\end{center}

\vspace{0.3cm} Note that the range of $\rho(G)$ (where $G \in {\cal
E}_{p.(2r)}$) will always contain at least $pr$ no. of pairwise
orthogonal product states, namely the states
$\frac{1}{\sqrt{L_{i^{\prime}k}}} \sum_{m = 1}^{L_{i^{\prime}k}}~
{\rm exp}~ \left(\frac{2 \pi i (m - 1) l}{L_{i^{\prime}k}}\right)
U_{i^{\prime}k}\left|a^{(i^{\prime}k)}_m\right\rangle \otimes
\frac{1}{\sqrt{2}} \left(\left|j_k\right\rangle -~ {\rm exp}~
\left(-\frac{2 \pi i l}{L_{i^{\prime}k}}\right)
\left|l_k\right\rangle\right)$ where $\sum_{i^{\prime} = 1}^{N_k}
L_{i^{\prime}k} = p$ and $k = 1, 2, \ldots, r$. The range of
$\rho(G)$ can also contain some other (possibly infinite in number)
product states if either (i) $L_{ik} = L_{i^{\prime}k}$ for
different $i$, $i^{\prime}$ in $\{1, 2, \ldots, N_k\}$, or (ii)
$\left\{a^{(ik)}_1, a^{(ik)}_2, \ldots, a^{(ik)}_{L_{ik}}\right\} =
\left\{a^{(ik^{\prime})}_1, a^{(ik^{\prime})}_2, \ldots,
a^{(ik^{\prime})}_{L_{ik^{\prime}}}\right\}$ for different $k,
k^{\prime} \in \{1, 2, \ldots, r\}$ (but for same $i$). All the
above-mentioned $pr$ no. of pairwise orthogonal product states are
reliably distinguishable by using local operations and classical
communication (LOCC).

Is there any ${\cal P}^S_{p.(2r)}$ such that $G \notin {\cal
E}_{p.(2r)}$? Yes, there are such perfect entangling matchings:
for $p = 3, r = 2$, there is (up to local permutations on the
labels of right hand and/or left hand sides) \emph{one} such $G$
which contains neither any cris-cross nor tally mark (see Figure
4). Higher the values of $p$ and/or $r$, higher will be the number
of such different $G$'s (not containing cris-crosses or tally
marks). From now on, we shall only consider those perfect
entangling matchings, none of which contains a cris-cross or tally
mark. Let $H \in {\cal P}^S_{p.(2r)} \backslash {\cal
E}_{p.(2r)}$. Is ${\rho}(H)$ a separable state in
$\mathbb{C}_{A}^{p}\otimes$ $\mathbb{C}_{B}^{2r}$? In order to
answer this question, we need to see whether there is any product
state (of $\mathbb{C}_{A}^{p}\otimes$ $\mathbb{C}_{B}^{2r}$)
within the range ($\mathcal{R}_{H}$, say) of $\rho(H)$. Also we
consider here the density matrix $\rho(H_{+})=\frac{1}{pr}\left( I
- \rho(H)\right)$, where $I$ is the $2pr \times 2pr$ identity
matrix. Let $\mathcal{R}_{H_{+}}$ be the range of $\rho(H_{+})$.
We have the following conjecture:

\vspace{0.2cm}
{\noindent {\bf Conjecture 2:} {\it Let $H \in
{\cal P}^S_{p.(2r)}\backslash {\cal E}_{p.(2r)}$ such that $H$
neither contains any cris-cross nor any tally mark. Then the range
$\mathcal{R}_{H}$ of $\rho(H)$ (the range $\mathcal{R}_{H_{+}}$ of
$\rho(H_{+})$) contains exactly $pr$ number of product states
$\left\vert {\psi}_{1}\right\rangle \otimes\left\vert
{\phi}_{1}\right\rangle $, $\left\vert {\psi}_{2}\right\rangle
\otimes\left\vert {\phi}_{2}\right\rangle $, $\ldots$, $\left\vert
{\psi}_{pr}\right\rangle \otimes\left\vert {\phi
}_{pr}\right\rangle $ of $\mathbb{C}_{A}^{p}\otimes$
$\mathbb{C}_{B}^{2r}$. Moreover, (i) all these product states are
pairwise orthogonal, (ii) all the
states $\left\vert {\psi}_{1}\right\rangle $, $\left\vert {\psi}%
_{2}\right\rangle $, $\ldots$, $\left\vert
{\psi}_{pr}\right\rangle $ are different but one can always have
at least one class of exactly $p$ of them all of which are
pairwise orthogonal, (iii) all the states $\left\vert {\phi
}_{1}\right\rangle $, $\left\vert {\phi}_{2}\right\rangle $,
$\ldots$, $\left\vert {\phi}_{pr}\right\rangle $ are different but
one can always have at least one class of exactly $2r$ of them all
of which are pairwise orthogonal, and (iv) all the states
$\left\vert {\psi}_{1}\right\rangle
\otimes\left\vert {\phi}_{1}\right\rangle $, $\left\vert {\psi}_{2}%
\right\rangle \otimes\left\vert {\phi}_{2}\right\rangle $,
$\ldots$,
$\left\vert {\psi}_{pr}\right\rangle \otimes\left\vert {\phi}_{pr}%
\right\rangle $ are reliably distinguishable by LOCC.} }

\vspace{0.3cm}
\begin{center}
\begin{figure}[h]
  \includegraphics[scale=0.4]{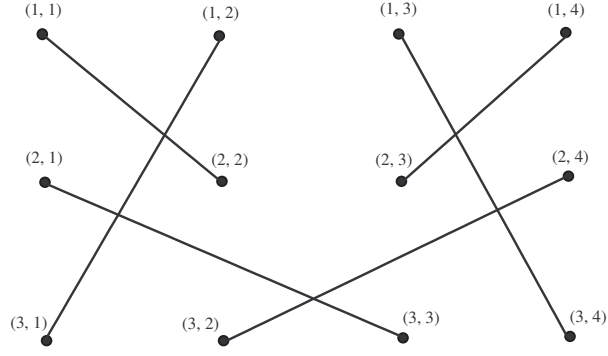}\\
  \caption{$G$ is the
representative perfect entangling matching on $12$ vertices $(1,
1)$, $(1, 2)$, $\ldots$ $(3, 4)$ such that $G \in {\cal
P}_{3.(2.2)}^S \backslash {\cal E}_{3.(2.2)}$.}
\label{figure-special-p-e-matching}
\end{figure}
\end{center}

\vspace{0.3cm} Validity of Conjecture 2 would directly show that
for any $H\in \mathcal{P}^S_{p.(2r)} \backslash
\mathcal{E}_{p.(2r)}$, which neither contains any cris-cross nor
any tally mark, $\rho(H)=\frac{1}{pr}\sum_{j = 1}^{pr} P\left[
\left\vert {\psi}_{j}\right\rangle \otimes\left\vert
{\phi}_{j}\right\rangle \right]$, and hence, $\rho(H)$ is
separable in $\mathbb{C}_{A}^{p}\otimes$ $\mathbb{C}_{B}^{2r}$. If
an $H \in \mathcal{P}^S_{p.(2r)} \backslash \mathcal{E}_{p.(2r)}$
contains some cris-crosses and/or tally marks, the rest part of
$H$ (eliminating out all these cris-crosses, tally marks) will be
again an element of $\mathcal{P}^S_{p^{\prime}.(2r^{\prime})}
\backslash \mathcal{E}_{p^{\prime}.(2r^{\prime})}$, for some
$p^{\prime}\leq p$ and $r^{\prime}\leq r$, such that this new
graph does not contain any cris-cross or tally mark. As
cris-crosses or tally marks always form separable density
matrices, therefore we see that for any $H \in
\mathcal{P}^S_{p.(2r)}$, $\rho(H)$ is separable in
$\mathbb{C}_{A}^{p}\otimes$ $\mathbb{C}_{B}^{2r}$, provided the
above-mentioned conjecture is true. Note that the validity of
Conjecture 1 automatically implies that for any $G \in {\cal
P}_{p.(2r)}$, $\rho(G)$ is separable if $G \in
\mathcal{P}^S_{p.(2r)} \backslash \mathcal{E}_{p.(2r)}$. However,
the statement in Conjecture 2 is much stronger than just saying
that $\rho(G)$ is separable if $G \in
 \mathcal{P}^S_{p.(2r)} \backslash \mathcal{E}_{p.(2r)}$.

\subsection{Proof of Theorem 5}

Let $\rho$ be a circulant density matrix of dimension $n=pq$. As
we already
mentioned in the introduction, we can write $\rho=\sum_{g\in\mathbb{Z}_{n}%
}f(g)\sigma(g)$ where $f$ is a complex-valued function. Obviously,
there will be some constraints imposed by the fact that $\rho$ is
positive semidefinite and hermitian. It is well-known that $\rho$
is diagonalized by the Fourier transform $FT(\mathbb{Z}_{n})$ over
$\mathbb{Z}_{n}$ \cite{da}: $[FT(\mathbb{Z}_{n})]_{j,k}=\exp(2\pi
ijk/n)$. The eigenvectors of $\rho$ are then the columns of
$(FT(\mathbb{Z}_{n}))^{\dagger}$. We prove the theorem in two
steps:\ (1)\ we prove that if $|\lambda\rangle$ is an eigenvector
of $\rho$ then $|\lambda\rangle=|a\rangle\otimes|b\rangle$, where
$|a\rangle \in\mathbb{C}_{A}^{p}$ and
$|b\rangle\in\mathbb{C}_{B}^{q}$, for any $p$ and $q$ such that
$n=pq$. (2) Then, for any chosen $p$ and $q$, we prove that
$\Delta(\rho)=\Delta(\rho^{\Gamma_{B}})$, where the partial
transpose is taken with respect to the standard orthonormal basis
$\{|ij\rangle :i=1,,,,.q;j=1,...,p\}\}$ of
$\mathbb{C}_{A}^{p}\otimes$ $\mathbb{C}_{B}^{q}$.

\bigskip

\noindent(1) Let $A$ be an $n\times n$ matrix which is
diagonalized by a unitary matrix $U$. We take $n=pq$. So
$UAU^{\dagger}=D$, where $D=$
diag$({\lambda}_{1},...,{\lambda}_{n})$, with respect to the
standard orthonormal basis $\{|\psi
_{1}\rangle,...,|\psi_{n}\rangle\}$. Thus
$A(U^{\dagger}|\psi_{i}\rangle) = \lambda_{i}(U^{\dagger}|\psi_{i}\rangle)$ for $i=1,2,...,n$, and $UAU^{\dagger}=\sum_{i=1}^{n}%
\lambda_{i}|\psi_{i}\rangle\langle\psi_{i}|$. Thus
$U^{\dagger}|\psi_i\rangle$ is an eigenvector of $A$ corresponding
to the eigenvalue ${\lambda}_i$. Also, $\langle\psi
_{i}|UU^{\dagger}|\psi_{j}\rangle=\langle\psi_{i}|\psi_{j}\rangle=\delta_{ij}%
$. Let $U$ be any $n \times n$ unitary matrix with its $(j, k)$-th
entry as $u_{jk}$ with respect to the orthonormal basis
$\{|\psi_1\rangle, |\psi_2\rangle, \ldots, |\psi_n\rangle\}$. Then $U^{\dagger}%
=(\omega_{jk})_{j,k=1}^{n}$ where $\omega_{jk}=u_{kj}^{\ast}$ for
all $j,k$.
Now $U^{\dagger}|\psi_{i}\rangle=\sum_{j=1}^{n}\omega_{ji}|\psi_{j}%
\rangle=\sum_{j=1}^{n}u_{ij}^{\ast}|\psi_{j}\rangle$, which is the
$i$-th column of $U^{\dagger}$. Assuming that $\rho$ is a
circulant matrix, with respect to
$\{|\psi_{1}\rangle,...,|\psi_{n}\rangle\}$, we have
\[
U=\frac{1}{\sqrt{n}}\left[
\begin{array}
[c]{ccccc}%
1 & 1 & 1 & \cdots & 1\\
1 & \omega & \omega^{2} & \cdots & \omega^{n-1}\\
1 & \omega^{2} & \omega^{4} & \cdots & \omega^{2(n-1)}\\
\vdots & \vdots & \vdots & \ddots & \vdots\\
1 & \omega^{n-1} & \omega^{2(n-1)} & \cdots & \omega^{(n-1)(n-1)}%
\end{array}
\right]  .
\]
The unitary matrix $U$ is the Fourier transform over
$\mathbb{Z}_{n}$
\cite{da}. We can see that%
\begin{align*}
\rho &  =\sum_{j=1}^{n}\lambda_{j}P[U^{\dagger}|\psi_{j}\rangle]\\
&  =\sum_{j=1}^{n}\lambda_{j}P\left[  \frac{1}{\sqrt{n}}\sum_{l=0}^{n-1}%
\exp\left(  -\frac{2\pi i(j-1)l}{n}\right)  |\psi_{l+1}\rangle\right] \\
&  =\sum_{j=1}^{n}\lambda_{j}P\left[  \frac{1}{\sqrt{n}}\sum_{a=0}^{p-1}%
\sum_{b=0}^{q-1}\exp\left(  -\frac{2\pi i(j-1)\left(  aq+b\right)  }%
{n}\right)  |a,b\rangle\right] \\
&  =\sum_{j=1}^{n}\lambda_{j}P\left[  \frac{1}{\sqrt{p}}\sum_{a=0}^{p-1}%
\exp\left(  -\frac{2\pi i(j-1)a}{p}\right)  |a\rangle\otimes\frac{1}{\sqrt{q}%
}\sum_{b=0}^{q-1}\exp\left(  -\frac{2\pi i(j-1)b}{pq}\right)
|b\rangle \right]  .
\end{align*}
It follows that $\rho$ is a separable density matrix provided that
$\lambda_{j}\geq0$ for $j=1,...,n$.

\noindent(2) The general form a circulant density matrix is%

\[
\rho=\left[
\begin{array}
[c]{cccc}%
a_{1} & a_{2} & \cdots & a_{n}\\
a_{n} & a_{1} & \cdots & a_{n-1}\\
\vdots & \vdots & \ddots & \vdots\\
a_{2} & a_{3} & \cdots & a_{1}%
\end{array}
\right]  .
\]
Since the matrix is symmetric, we have
\begin{equation}
a_{1} = a_{1}^{\ast},a_{2}=a_{n}^{\ast},a_{3}=a_{n-1}^{\ast},...,a_{l}=a_{n-l+2}^{\ast}. \label{sy}%
\end{equation}
Consider $\rho$ as a block-matrix with $p^{2}$ blocks, each block
being a
$q\times q$ matrix:%
\[
\rho=\left[
\begin{array}
[c]{cccc}%
A_{1,1} & A_{1,2} & \cdots & A_{1,p}\\
A_{2,1} & A_{2,2} & \cdots & A_{2,p}\\
\vdots & \vdots & \ddots & \vdots\\
A_{p,1} & A_{p,2} & \cdots & A_{p,p}%
\end{array}
\right]  .
\]
Consider the block $A_{1,m+1}$. Then $[A_{1,m+1}]_{1,1}=a_{mq+1}$.
Let
$l=mq+1$. Then%
\[
A_{1,m+1}=\left[
\begin{array}
[c]{cccc}%
a_{l} & a_{l+1} & \cdots & a_{l+q-1}\\
a_{l-1} & a_{l} & \cdots & a_{l+q-2}\\
\vdots & \vdots & \ddots & \vdots\\
a_{l-q+1} &  & \cdots &
\end{array}
\right]  .
\]
Now, consider the block $A_{1,p-m+1}$. Then $[A_{1,p-m+1}]_{1,1}%
=a_{n-l+2}=a_{n+1-mq}$. Then%
\[
A_{1,n-m+1}=\left[
\begin{array}
[c]{cccc}%
a_{n-l+2} & a_{n-l+3} & \cdots & a_{n-l+q+1}\\
a_{n-l+1} & a_{n-l+2} & \cdots & a_{n-l+q}\\
\vdots & \vdots & \ddots & \vdots\\
a_{n-l-q+3} &  & \cdots &
\end{array}
\right]
\]
Applying the condition expressed in Equation (\ref{sy}), one can
verify that
\[
A_{1,m+1}=A_{1,n-m+1}^{\dagger}.
\]
This argument extends to all blocks of the $i$-th block-row of
$\rho$. For
example, the first block-row of $\rho$ is then of the form%
\[%
\begin{array}
[c]{ccccccccc}%
A_{1,1}\left(  =A_{1,1}^{\dagger}\right)  & A_{1,2} & A_{1,3} &
\cdots & A_{1,p/2+1} &
A_{1,p/2+1}^{\dagger} & \cdots & A_{1,3}^{\dagger} & A_{1,2}^{\dagger}%
\end{array}
\]
if $p$ is even, and
\[%
\begin{array}
[c]{cccccccc}%
A_{1,1}\left(  =A_{1,1}^{\dagger}\right)  & A_{1,2} & A_{1,3} & \cdots & A_{1,(p+1)/2}%
\left(  =A_{1,(p+1)/2}^{\dagger}\right)  & \cdots &
A_{1,3}^{\dagger} &
A_{1,2}^{\dagger}%
\end{array}
\]
if $p$ is odd. It is then clear that $\Delta(\rho)=\Delta\left( \rho
^{\Gamma_{B}}\right)  $, that is the the row sums of $\rho$ are
invariant under the partial transpose. It should be noted here that
each element of $\Delta(\rho)$ (as well as of
$\Delta({\rho}^{{\Gamma}_B})$) is real due to equation (2.15).

The same reasoning applies to the second part of the theorem. The
only difference is that
$\rho=\sum_{g\in\mathbb{Z}_{2}^{n}}f(g)\sigma(g)$ is diagonalized
by the Hadamard matrices of Sylvester type, $H^{n}=H^{n-1}\otimes
H$, where $H$ is the $2\times2$ Hadamard matrix \cite{terras}.

\section{Open problems}

In this paper we have studied the separability of a class of states
associated with the combinatorial laplacians of graphs. The graphs
for these states compactly encodes information about their bipartite
\ entanglement. We have shown that invariance of the degree matrices
under partial transposition gives, in many cases, significant
information about the separability of the states. Now the
Peres-Horodecki partial transposition condition (known as the {\it
PPT criterion}) is only a necessary condition (in general) for
separability of any bipartite density matrices \cite{P}, \cite{H}.
In fact, all the {\it practical} separability conditions, available
so far, are either necessary or sufficient for general bipartite
density matrices (see, for example, \cite{l}). The degree condition,
described in this paper, is of course weaker than the PPT criterion,
as not all bipartite density matrices (not even the separable ones)
can be described as density matrices generated from graphs.
Nevertheless the validity of Conjecture 1 (together with Theorem 2)
would imply that the degree condition is {\it both} necessary as
well as sufficient for some particular classes of bipartite density
matrices {\it irrespective} of the dimension of the system. In the
quest for resolving the separability problem with the help of
practical necessary-sufficient conditions ({\it i.e.}, conditions,
each of which is both necessary as well as sufficient), one possible
way would be to find out the set of all possible independent but
practical necessary-sufficient conditions each of which decides the
separability problem of a {\it maximal} set of bipartite density
matrices in such a way that the collection of these later sets would
comprise the entire set of bipartite density matrices. The present
work is one step forward in that direction. The following points are
open for further investigation:

\bigskip

\noindent\textbf{A. (Partial transposition as a local
permutation)} It is not
difficult to see that for any graph on $n=pq$ vertices $v_{1}=u_{1}%
w_{1},...,v_{pq}=u_{p}w_{q}$, if $\Delta(G)=\Delta\left(  G^{\Gamma_{B}%
}\right)  $ then there is a permutation matrix $P$ on the labels
$w_{1},...,w_{q}$ such that
\[%
\begin{tabular}
[c]{lll}%
$\Delta\left(  G^{\Gamma_{B}}\right)  =\left(  I\otimes P\right)
\Delta(G)\left(  I\otimes P^{-1}\right)  $ & and &
$M(G^{\Gamma_{B}})=\left(
I\otimes P\right)  M(G)\left(  I\otimes P^{-1}\right)  .$%
\end{tabular}
\]
This says that if the degree condition is satisfied then the
operation of partial transposition is nothing but a local
permutation. Note that this is, in general, false for the case of
any given bipartite separable density matrix. The relation between
separability of density matrices of graphs and isomorphism remains
to be studied.

\bigskip

\noindent\textbf{B.} \textbf{(Structure of bipartite Hilbert
spaces) }The validity of of Conjecture 6 can be traced back to
basic problems in the structure of any bipartite Hilbert space.
Given a subspace $\mathcal{S}$ of dimension $d$ of $\mathcal{H}$,
what are the necessary and sufficient conditions under which at
least one of the following situations hold good?

\begin{enumerate}
\item $\mathcal{S}$ contains at least one linearly independent product state;

\item $\mathcal{S}$ contains only $d^{\prime}$ linearly independent product states,
where $d^{\prime}<d$;

\item $\mathcal{S}$ contains more than $d$ product states (in which case they must be linearly dependent);

\item $\mathcal{S}$ contains contains exactly $d$ linearly independent product states
and these are pairwise orthogonal;

\item $\mathcal{S}$ contains only $d^{\prime}$, where
$d^{\prime}\leq d$, pairwise orthogonal product states that one
can extend to a full orthogonal product basis of $\mathcal{H}$,
\emph{etc}.
\end{enumerate}

\bigskip

\noindent\textbf{C.} \textbf{(Multiparty entanglement) }As a
generalization of our result to density matrices of graphs having
multiple labels on their vertices, we expect that if $G$ is a
graph on $n=p_{1}p_{2}\ldots p_{m}$
vertices%
\[%
\begin{tabular}
[c]{lll}%
$v_{i}=u_{1s_{1}^{(1)}}u_{2s_{i}^{(2)}}\ldots u_{ms_{i}^{(m)}},$ &
& where
$s_{i}^{(j)}\in\{1,...,p_{j}\}$ for $j=1,...,m$ and $i=1,...,n,$%
\end{tabular}
\
\]
then $\rho(G)$ is a separable density matrix in $\mathbb{C}_{A_{1}%
...A_{j-1}A_{j+1}...A_{m}}^{p_{1}p_{2}\ldots p_{j-1}p_{j+1}...p_{m}}%
\otimes\mathbb{C}_{A_{j}}^{p_{j}}$ if and only if
$\Delta(G)=\Delta (G^{\Gamma_{A_{j}}})$. Moreover, we expect that
$\rho(G)$ is a completely
separable density matrix in $\mathbb{C}_{A_{1}}^{p_{1}}\otimes\mathbb{C}%
_{A_{2}}^{p_{2}}\otimes\mathbb{\cdots}\otimes\mathbb{C}_{A_{m}}^{p_{m}}$
if and only if $\Delta(G)=\Delta(G^{\Gamma_{A_{j}}})$ for
$j=1,...,m$.

\bigskip

\noindent\textbf{Acknowledgement} SLB currently holds a Royal
Society --- Wolfson Research Merit Award. SG, SS and RCW
acknowledge the support received from the EPSRC. Part of of this
work has been carried out while SS was at The Caesarea Rothschild
Institute, Israel.

\end{document}